\documentclass[twocolumn,prl,showpacs,preprintnumbers,amsmath,amssymb]{revtex4}

\usepackage{graphicx}
\usepackage{dcolumn}
\usepackage{bm}

\begin{document}


            
\title{Comment on ``Field-Enhanced Diamagnetism in the Pseudogap State of the Cuprate Bi$_2$Sr$_2$CaCu$_2$O$_{8+\delta}$
Superconductor in an Intense Magnetic Field''}

\author{Luc\'ia Cabo}  
\author{Jes\'us Mosqueira}
\author{F\'elix Vidal}
\affiliation{LBTS, Universidade de Santiago de Compostela, E-15782, Spain}
\maketitle

In a recent Letter\cite{Wang}, Wang \textit{et al}.~claim that ``the magnetization
results above $T_C$ distinguish $M$ from conventional amplitude fluctuations and strongly support the vortex
scenario for the loss of phase coherence at $T_C$.''
However, we will present here some examples for $T>T_C$ that show that the data of Ref.\cite{Wang} may be explained on the grounds of the conventional Ginzburg-Landau (GL) scenario. We have checked that this conclusion applies also to \textit{both} the temperature and the magnetic field dependende of $M$ below $T_C$.

Note first that the absence on non-local electrodynamic effects on the fluctuation magnetization $M'$ measured in Ref.\cite{Wang} just confirms earlier results (see, e.g., Refs.\cite{Lee,Carballeira00}). A similar absence was also observed in dirty low-$T_C$ superconductors\cite{Mosqueira} and therefore, contrary to the claims of Wang \textit{et al}., it does not provide a ``first evidence" of unconventional fluctuations.

To estimate $M'(T,H)$ above but not too close to $T_C$ we may use the GL theory with Gaussian fluctuations (GGL approach) in the 2D limit\cite{Carballeira},
  
\begin{eqnarray}
\nonumber
M'=-f\frac{k_BTN}{\phi_0s}\left[-\frac{\epsilon^c}{2h}\psi\left(\frac{h+\epsilon^c}{2h}\right)-\mathrm{ln}\Gamma\left(\frac{h+\epsilon}{2h}\right)\right.
\\
\left.+\mathrm{ln}\Gamma\left(\frac{h+\epsilon^c}{2h}\right)+\frac{\epsilon}{2h}\psi\left(\frac{h+\epsilon}{2h}\right)+\frac{\epsilon^c-\epsilon}{2h}\right]. 
\end{eqnarray}
The notation is the same as in Ref.\cite{Carballeira}, with $\epsilon^c\equiv\ln(T^C/T_C)\approx0.55$ as the \textit{total energy} cutoff constant.\cite{Vidal} The temperature above which all fluctuations vanish is then $T^C\approx1.7\; T_C$.\cite{Vidal} 
The constant $f$ accounts for possible effective superconducting fractions lower than 1 and sample misalignments. 

\begin{figure}[t]
\includegraphics[width=.5\textwidth]{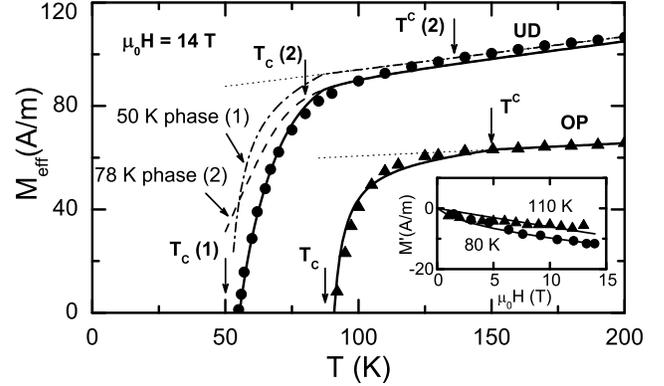}
\caption{Comparison between the Wang \textit{et al.} magnetization measurements above $T_C$, $M_{\rm eff}$ (solid data points), and the GGL approach (solid lines).}
\end{figure}

For the optimally doped (OP) crystal, the solid line in Fig.~1 was obtained by fitting Eq.~(1) plus the corresponding background magnetization  (dotted line) to the data of Wang \textit{et al}.~[their Fig.~1(b)].
We used $\mu_0H_{C2}(0)\approx300$ T\cite{Bi2212}, and $f$ is the only free parameter.   
This leads to $f\approx0.5$, a value quite reasonable for this kind of compound.\cite{Kogan} 
The agreement includes the predicted $T^C\approx1.7T_C\approx150$~K. A good agreement is also found for the underdoped crystal (UD) by adding to the magnetization of the low-$T_C$ phase [dot-dashed line, with $T_C(1)=50$ K] a contribution from the minority phase with $T_C(2)\approx78$ K (dashed line), estimated by using the Te$\check{\rm{s}}$anovi$\acute{\rm{c}}$ \textit{et al.}~expression for $M'$, valid in the critical region around $T_C$\cite{Tesanovic}. Note that $T^C(2)\approx1.7\; T_C(2)\approx135$~K is again close to the measured $T_{\rm{onset}}$. 

Not too close to $T_C$, the $H$ dependence of $M'$ may also be easily explained in terms of the GGL approach (see the inset in Fig.~1). Near $T_C$, one must take into account not only the uncertainties on $T_C$, $H_{C2}(0)$ and $f$, but mainly the $T_C$ inhomogeneities,\cite{Lucia} these last explicitly recognized by Wang \textit{et al}. In fact, measuring samples deeply affected by extrinsic inhomogeneities is not an enlightening exercise, \textit{independently of the field amplitude used}.
Nevertheless, we have also checked that the data below $T_C$, including the seemingly ``anomalous'' $H_{C2}(T)$ behavior and the ``striking'' nonlinear $M(H)_T$ curves, are accounted by the GL approach with conventional vortex fluctuations.\cite{Kogan,Tesanovic,Bulaevski,Welp} So, when correctly analyzed the magnetization data of Wang \textit{et al}.~directly contradict their own proposals about unconventional (non GL) Meissner transition at $T_C$ in cuprates.


\begin{references}
\bibitem{Wang}Y. Wang \textit{et al.}, Phys. Rev. Lett. {\bf 95}, 247002 (2005).
\bibitem{Lee}W.C. Lee, R.A. Klemm, and D.C. Johnston, Phys. Rev. Lett. {\bf 63}, 1012 (1989).
\bibitem{Carballeira00}C. Carballeira {\it et al.}, Phys. Rev. Lett. {\bf 84}, 3157 (2000).
\bibitem{Mosqueira}J. Mosqueira \textit{et al.}, Phys. Rev. Lett. {\bf 87}, 167009 (2001).   
\bibitem{Carballeira}C. Carballeira \textit{et al.}, Physica C {\bf 384}, 185 (2003). 
\bibitem{Vidal}F. Vidal \textit{et al.}, Europhys. Lett. {\bf 59}, 754 (2002).
\bibitem{Bi2212}Q. Li \textit{et al.}, Phys, Rev. B {\bf 46}, 3195 (1992).
\bibitem{Kogan}V.L. Kogan {\it et al.}, Phys. Rev. Lett. {\bf 70}, 1870 (1993).
\bibitem{Tesanovic} Z. Te$\check{\rm{s}}$anovi$\acute{\rm{c}}$ \textit{et al.}, Phys. Rev. Lett. \textbf{69}, 3563 (1992).
\bibitem{Lucia} L. Cabo {\it et al.}, Phys. Rev. B {\bf 73}, 184520 (2006).
\bibitem{Bulaevski} L.N. Bulaevskii, M. Ledvij, and V.G. Kogan, Phys. Rev. Lett. {\bf 68}, 3773 (1992).
\bibitem{Welp}U. Welp {\it et al.}, Phys. Rev. Lett. {\bf 67}, 3180 (1991).


\end{references}
\end{document}